\documentclass[sigconf]{acmart}

\usepackage{graphicx}
\usepackage{multirow}

\begin{document}

\title{\textit{EVA}: Generating Emotional Behavior of Virtual Agents using Expressive Features of Gait and Gaze}

\author{Tanmay Randhavane}
\affiliation{%
  \institution{University of North Carolina}
  \city{Chapel Hill}
  \country{USA}}
\email{tanmay@cs.unc.edu}

\author{Aniket Bera}
\affiliation{%
  \institution{University of North Carolina}
  \city{Chapel Hill}
  \country{USA}}
\email{ab@cs.unc.edu}

\author{Kyra Kapsaskis}
\affiliation{%
  \institution{University of North Carolina}
  \city{Chapel Hill}
  \country{USA}}
\email{kyrakaps@gmail.com}

\author{Rahul Sheth}
\affiliation{%
  \institution{Snap Inc.}
  \city{Santa Monica}
  \country{USA}}
\email{rahul@snap.com}

\author{Kurt Gray}
\affiliation{%
  \institution{University of North Carolina}
  \city{Chapel Hill}
  \country{USA}}
\email{kurtgray@unc.edu}

\author{Dinesh Manocha}
\affiliation{%
  \institution{University of Maryland}
  \city{College Park}
  \country{USA}}
\email{dm@cs.umd.edu}


\begin{abstract}
We present a novel, real-time algorithm, \textit{EVA}, for generating virtual agents with various perceived emotions. Our approach is based on using \textit{Expressive Features} of gaze and gait to convey emotions corresponding to happy, sad, angry, or neutral. We precompute a data-driven mapping between gaits and their perceived emotions. EVA uses this gait emotion association at runtime to generate appropriate walking styles in terms of gaits and gaze. Using the EVA algorithm, we can simulate gaits and gazing behaviors of hundreds of virtual agents in real-time with known emotional characteristics. We have evaluated the benefits in different multi-agent VR simulation environments. Our studies suggest that the use of expressive features corresponding to gait and gaze can considerably increase the sense of presence in scenarios with multiple virtual agents.
\end{abstract}


\maketitle

\section{Introduction}

\begin{figure}[t!]
    \centering
    \includegraphics[width=\linewidth]{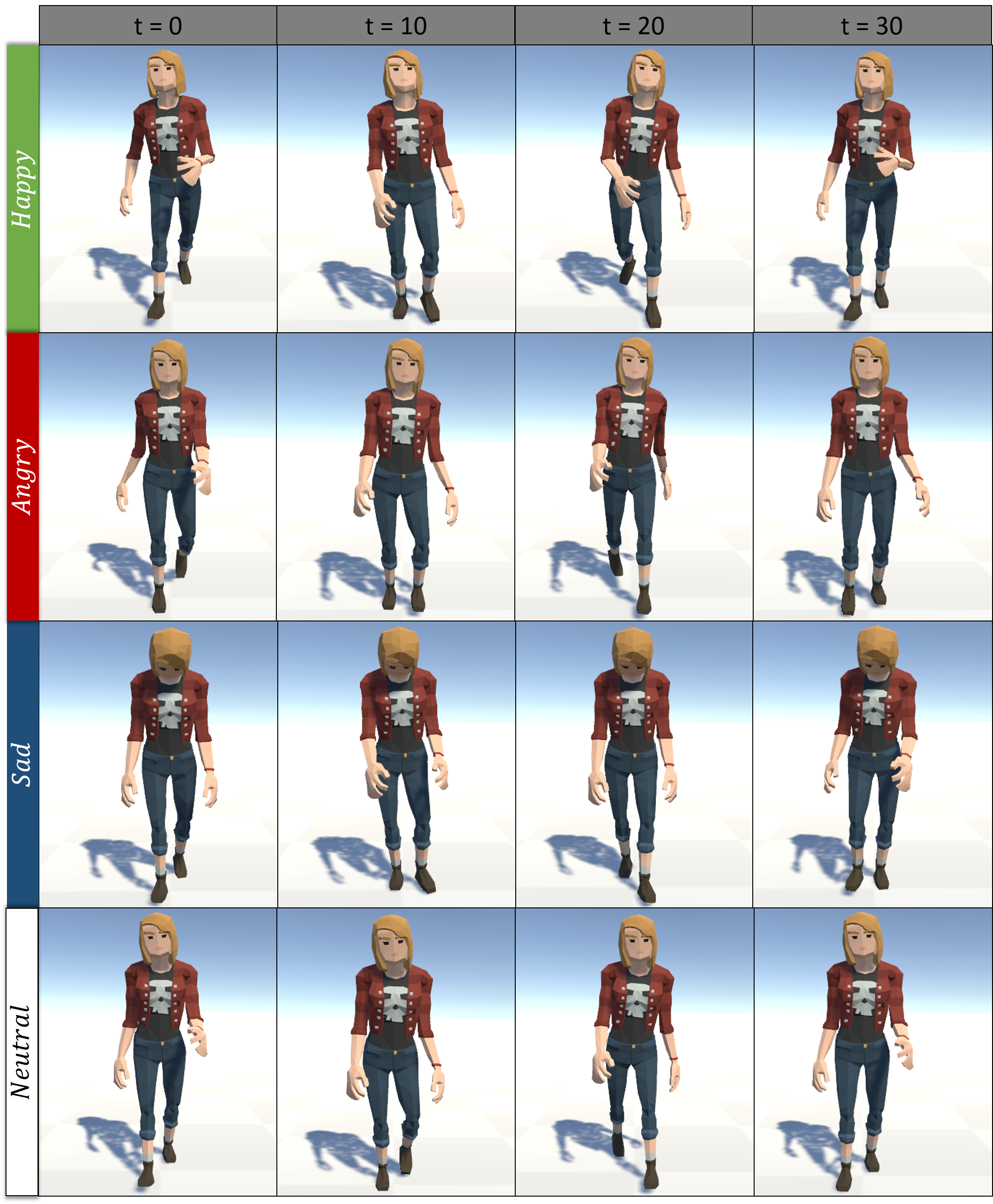}
    \vspace{-18pt}
    \caption{We present a novel, real-time algorithm, \textit{EVA}, for generating virtual agents with various emotions. We compute a novel data-driven gait emotion association and combine them with gazing behaviors to create virtual agents with different emotions. To simulate gazing, we use expressive features such as direct (Happy, Angry) vs. averted gazing (Sad) and neck flexion (Sad) vs. extension (Happy). To simulate gaits, we use expressive features such as body expansion (Happy), compact posture (Sad), rapid and increased movements (Angry, Happy), and slow movements (Sad).}
    \label{fig:cover}
    \vspace{-15pt}
\end{figure}

In recent years, interest in developing intelligent virtual agents has increased because of the applications in many areas including gaming and social VR~\cite{ferstl2018perceptual},  evacuation simulations~\cite{liu2018perception}, therapy and rehabilitation environments~\cite{rivas2015detecting}, training, etc. These applications require simulation of virtual agents that have personalities, moods, emotions and exhibit human-like behaviors. Social robots, too, require human-like appearance and behaviors to elicit more empathy from humans interacting with them~\cite{riek2009anthropomorphism}. In the physical world, humans communicate emotions and moods using speech, gestures, and facial expressions. Non-verbal expressions, including body posture and movements, are also important and used for social interactions. Virtual agents and robots that can communicate using both verbal and non-verbal channels are, therefore, important for human-computer and human-robot interactions.

Emotions influence our interactions with the environment and other agents~\cite{barrett2017}. The perception of someone's emotional state helps to define how we view their personality and how we react to their actions~\cite{ConfirmationBiasNickerson1998}. Therefore, it is crucial to incorporate the emotional context in the simulation of intelligent virtual agents. In this paper, we mainly focus on conveying emotions of virtual agents based on gait and gaze.

Psychology literature refers to the features that are used to communicate emotional states through nonverbal movements as \textit{Expressive Features}~\cite{Riggio2017}. These features include facial expressions, body postures, and movements~\cite{kleinsmith2013affective}. During walking, an observer uses the style of walking (i.e., gait) to assist the perception of emotions~\cite{monteparegaitinformation}. Gait features such as the extent of arm swings, the foot landing force, length of the strides, and erectness of the posture have been shown to assist in the identification of emotions such as sadness, anger, happiness, and pride~\cite{monteparegaitinformation}. For example, faster walking combined with high arm swings is perceived as happy, and a collapsed upper body with less movement activity is considered as sad~\cite{kleinsmith2013affective}.

In addition to gaits, gazing behaviors also affect the perception and expression of emotions~\cite{gallup2014influence}. As one of the critical components of non-verbal communication~\cite{bailenson2005transformed}, gaze increases the plausibility of embodied conversational agents (ECA)~\cite{peters2005model}. A direct versus an averted gaze also changes the perceptions of emotions; direct gaze is associated with anger, whereas averted gaze is associated with sadness~\cite{adams2003perceived}. In this paper, we restrict ourselves to expressive features of gait and gaze features.

\textbf{Main Results}: We present two novel contributions:

\noindent \textbf{1)} A data-driven metric between gaits and perceived emotions based on a user study. We chose \textit{happy, angry, sad}, and \textit{neutral} emotions because they are known to persist for a prolonged interval of time and are more apparent during walking~\cite{ma2006motion}. Using the continuous space representation of emotions~\cite{kleinsmith2005grounding,ekman1967head,mehrabian1980basic,mehrabian1996pleasure}, we can combine these four basic emotions to represent all other emotions of virtual agents~\cite{mikels2005emotional,morris1995observations}. This gait emotion association can also be used to predict the perceived emotion of any new input gait with an accuracy of $70.04\%$.


\noindent \textbf{2) EVA}: A novel, data-driven approach for generating emotions in virtual agents, \textit{EVA}, using the expressive features of gait and gaze (Figure~\ref{fig:cover}). EVA algorithm uses the gait emotion association to generate gaits corresponding to different emotions. We combine these gaits with gazing behaviors and create virtual agents that are perceived as experiencing \textit{happy, angry, sad}, or \textit{neutral} emotions.\footnote{Please refer to the accompanying video for visual results.}

We validate our algorithms using standard metrics of co-presence \cite{garau2005responses}. The questionnaire measures the level of presence achieved by a real user immersed in a virtual environment, based on the emotional state of virtual agents generated using EVA. Our results indicate that in scenarios with a high number of virtual agents, expressive features can increase the sense of presence felt by the users in the virtual environment. Our results also indicate that both gait and gazing features contribute to the perceptions of emotions in virtual agents. We observe that  EVA simulates virtual agents with desired emotions with an accuracy of $70.83\%$, and achieves $100\%$ accuracy in terms of simulating emotion corresponding to sadness.

This paper is organized as follows. In Section 2, we present related work on expressive features and emotions of virtual agents. In Section 3, we give a brief overview of our virtual agent generation algorithm and present the details of the gait emotion association. 
In Section 4, we present the details of our EVA algorithm, which uses the expressive features to generate emotions in virtual agents.  We present the details of our user evaluation in Section 5.

\section{Related Work}
\label{sec:RelatedWork}
In this section, we give a brief overview of related work on expressive features and emotions of virtual agents.

\subsection{Expressive Features of Virtual Agents}
Emotional states can be communicated through expressive features such as nonverbal movements and gestures, including facial expressions~\cite{Riggio2017}. Ferstl et al.~\shortcite{ferstl2018perceptual} studied the effect of the manipulation of facial features for trait portrayal in virtual agents. Previous research has used trajectories to convey dominance~\cite{randhavane2018pedestrian}, approachability~\cite{randhavane2017f2fcrowds}, group entitativity~\cite{bera2018data}, and personalities~\cite{bera2017aggressive}. Biometric data can also be used for predicting the emotions of individuals~\cite{yates2017arousal}. In this paper, we focus on the expressive features of gait and gaze movements. Different methods have been proposed to simulate the gait and gaze of virtual agents to convey a variety of behaviors in virtual agents. Narang et al.~\shortcite{narang2017motion} studied self-recognition in virtual agents with walking gaits and avatars of real humans. Gaits have been used to convey friendliness~\cite{randhavane2019FVA} and dominance~\cite{randhavane2019modeling} of virtual agents. Gazing has been a focus of attention in many VR studies too~\cite{vinayagamoorthy2004eye,pedvr}. Loth et al.~\shortcite{loth2018accuracy} studied how users perceived the gaze of virtual agents. Bailenson et al.~\shortcite{bailenson2005transformed} investigated the relationship between gazing and personal space. In this paper, we combine both gait and gaze behaviors to convey emotions in virtual agents.

\subsection{Emotions of Virtual Agents}
Affective computing has been applied to improve the simulation of many human-like behaviors. Jaques et al.~\shortcite{jaques2016understanding} investigated the design of intelligent agents to promote bonding effectively.  Liebold et al.~\cite{liebold2016cognitive} discuss how humans understand and process emotions in virtual environments. Lee et al.~\cite{lee2006feeling} present an emotion model for virtual agents that deals with the resolution to ambivalence, in which two emotions conflict. Pelczer et al.~\cite{pelczer2007expressions} present a method to evaluate the accuracy of the recognition of expressions of emotions in virtual agents.  

Many approaches have been proposed to simulate emotions in virtual agents using verbal communication~\cite{sohn2018emotionally,Chowanda2016computational}, facial expressions~\cite{ferstl2018perceptual,Chowanda2016computational,badler2002representing,tinwell2011facial,liu2005emotion,beer2009emotion}, gaze~\cite{lance2008model}, gaits~\cite{mchugh2010perceiving}, and gestures~\cite{pelachaud2009studies}. McHugh et al.~\shortcite{mchugh2010perceiving} studied the role of dynamic body postures on the perception of emotion in crowded scenes. Clavel et al.~\shortcite{clavel2009combining} evaluated how both face and posture modalities affect the perceptions of emotions in virtual characters. Lance and Marsella~\shortcite{lance2008model} used a generative model of expressive gaze for emotional expression in virtual agents. Many generalized virtual agent modeling frameworks have been developed (e.g., SmartBody~\shortcite{thiebaux2008smartbody}) which model the animation, navigation, and behavior of virtual agents. Vinayagamoorthy et al.~\cite{vinayagamoorthy2006building} describe models of individual characters' emotions and personalities, models of interpersonal behaviors, and methods for generating emotions using facial expressions.  Liebold et al.~\cite{liebold2013multimodal} observe that multimodal (e.g., a combination of verbal cues and facial expression) expressions of emotions yield the highest recognition rates. Inspired by these approaches, we use a perception study (for gait features) and psychological characterization (for gaze features) to simulate emotions of virtual agents.

\section{Gait Emotion Association Metric}
In this section, we provide an overview of our algorithm for generating emotions of virtual agents. We also describe the gait emotion metric. This metric is used at runtime to generate gaits corresponding to different emotions.

\begin{figure}[t]
  \centering
  \includegraphics[width =\linewidth]{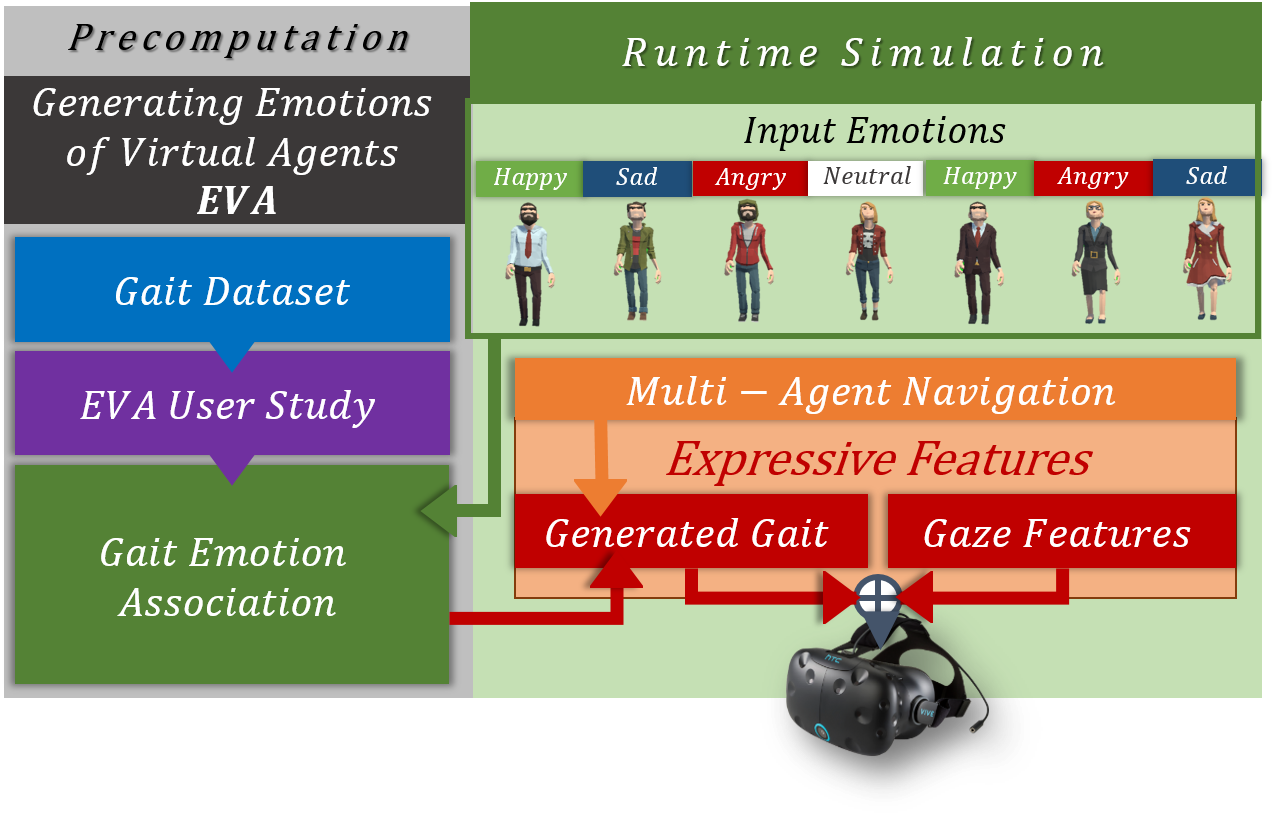}
  \vspace{-28pt}
  \caption{\textbf{Overview:} Our virtual agent generation algorithm consists of two parts: (1) offline computation of gait emotion association and (2) real-time generation of virtual agents (EVA) with desired emotional characteristics using expressive features. We obtain emotion labels for a set of motion-captured gaits with a user study and model the gait emotion association metric using these labels. At runtime, we take desired emotions for each virtual agent as input and generate expressive features corresponding to the desired emotions.}
  \vspace{-15pt}
  \label{fig:overview}
\end{figure}

\subsection{Overview}
Our EVA algorithm (Figure~\ref{fig:overview}) consists of two parts: (1) offline computation of an association between gaits and emotions, and (2) real-time generation of virtual agents with desired emotional characteristics using expressive features. We obtain emotion labels for a set of motion-captured gaits with a user study. We use these labels and formulate a data-driven association between the gaits and emotions. We can also use this association to predict the perceived emotion of any input gaits. At runtime, we take desired emotions for each virtual agent as inputs and use the gait emotion association to generate gaits corresponding to the desired emotions. We also combine our method with a multi-agent navigation algorithm (e.g., RVO~\cite{van2008reciprocal}) for collision-free navigation of the agents in the virtual environment. We combine gaze features according to the desired emotion for each virtual agent. 

\subsection{Gait Dataset}\label{sec:datasets}
We used a combination of publicly available datasets of walking gaits to formulate our gait emotion association metric: 

\noindent $\bullet$ \textbf{BML}~\cite{ma2006motion}: This motion-captured dataset contains $120$ different gaits with four different walking styles $30$ subjects ($15$ male and $15$ female).

\noindent $\bullet$ \textbf{CMU}~\shortcite{CMUGait}: This  motion-captured dataset contains $49$ gaits obtained from subjects walking with different styles.

\noindent $\bullet$ \textbf{EWalk}: This dataset contains gaits extracted from $94$ RGB videos using state-of-the-art 3D pose estimation~\cite{dabral2018learning}.

\noindent $\bullet$ \textbf{Human3.6M}~\cite{ionescu2013human3}: This motion-captured dataset contains $14$ gaits acquired by recording the performance of five female and six male subjects walking.

\noindent $\bullet$ \textbf{ICT}~\cite{narang2017motion}: This motion-captured dataset contains $24$ gaits obtained from videos of $24$ subjects.

\noindent $\bullet$ \textbf{SIG}~\cite{xia2015realtime}: This is a synthetic dataset of $41$ gaits generated using local mixtures of autoregressive models.

\begin{figure}[t]
  \centering
  \includegraphics[width =\linewidth]{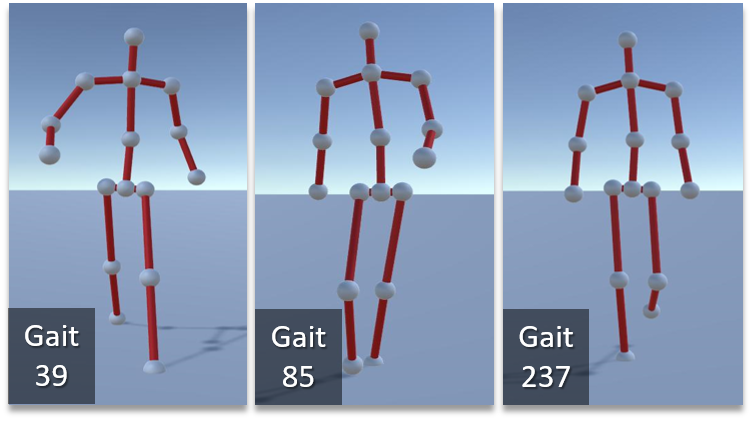}
  \vspace{-18pt}
  \caption{\textbf{Gait Visualization:} We presented $342$ gaits visualized from the viewpoint of a camera situated in front of the mesh. }
  \vspace{-18pt}
  \label{fig:gaitVideo}
\end{figure}

We visualized each gait using a skeleton mesh (Figure~\ref{fig:gaitVideo}). We rendered the gait visualizations from the viewpoint of a camera situated in front of the skeleton mesh and generated a total of $342$ gait visualizations.


\subsection{EVA User Study}
To compute our gait emotion association metric, we showed the $342$ gait visualizations to participants in a web-based user study. Based on the participant responses, we obtained emotion labels for each video.

\subsubsection{Experiment Details}
We recruited $688$ participants ($279$ female, $406$ male, $\bar{age} = 34.8$) from Amazon Mechanical Turk. We presented $10$ randomly chosen videos to each participant. After watching each video, the participant answered whether he/she perceived the gait in the video as happy, angry, sad, or neutral on a 5-point Likert scale ranging from \textit{strongly disagree} to \textit{strongly agree}. For each video, we obtained a minimum of $10$ participant responses.

\subsubsection{Results}
We converted the participant responses to an integer scale from $[1,5]$, where $1$ corresponds to \textit{strongly disagree}, and $5$ corresponds to \textit{strongly agree}. For each emotion $e$, we obtained the mean of all participant responses ($r^{e}_{i, j}$) for each gait $\textbf{G}_i$ in the datasets:
\begin{eqnarray}
    r^{e}_i = \frac{\sum_{j=1}^{n_p} r^{e}_{i,j}}{n_p},
\end{eqnarray}
where $j$ is the participant id and $n_p$ is the number of participant responses collected.

We computed the correlation between participants' responses to the questions relating to the four emotions (Table~\ref{tab:correl}). A correlation value closer to $1$ indicates that the two variables are positively correlated and a correlation value closer to $-1$ indicates that the two variables are negatively correlated. A correlation value closer to $0$ indicates that two variables are uncorrelated. As expected, \textit{happy} and \textit{sad} are negatively correlated and \textit{neutral} is uncorrelated with the other emotions.

\begin{table}[t]
\centering
\begin{tabular}{|c|c|c|c|c|}
\hline
        & Happy  & Angry  & Sad    & Neutral \\ \hline
Happy   & 1.000  & -0.268 & -0.775 & -0.175  \\ \hline
Angry   & -0.268 & 1.000  & -0.086 & -0.058  \\ \hline
Sad     & -0.775 & -0.086 & 1.000  & -0.036  \\ \hline
Neutral & -0.175 & -0.058 & -0.036 & 1.000   \\ \hline
\end{tabular}
\caption{Correlation Between Emotion Responses: We present the correlation between participants' responses to questions relating to the four emotions.}
\vspace{-25pt}
\label{tab:correl} 
\end{table}

\subsection{Gait Emotion Association}
We used the participant responses to obtain emotion labels for the gaits in the gait datasets. We obtained the emotion label $e_i$ for a gait $\textbf{G}_i$ as  $e_i = e \mid r^{e}_i > \theta$,
where $\theta = 3.5$ is an experimentally determined threshold for emotion perception.

If $r^{e}_i < \theta$ for all $4$ emotions, then the corresponding gait $G_i$ is not emotionally expressive and we do not use it for metric computation ($21$ gaits). If there are multiple emotions with average participant responses greater than $ r^{e}_i > \theta$, we choose the emotion with the highest average participant response as the label $e_i$ for the gait $G_i$. 

Using these gaits $G_i$ and emotion labels $e_i$, we define a function $GEA(e)$ that returns a set of gaits $G_i$ for which the emotion label $e_i$ matches with the emotion $e$ as $GEA(e) = \{G_i | e_i = e \}$. We refer to this function as our novel gait emotion association metric.

\begin{figure}[t]
  \centering
  \includegraphics[width =\linewidth]{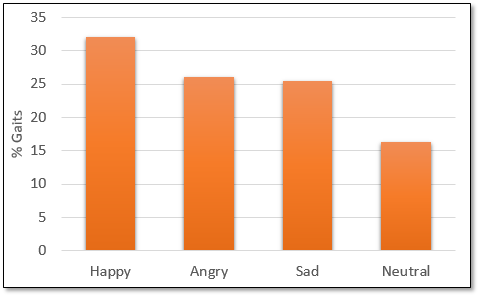}
  \vspace{-18pt}
  \caption{\textbf{Emotion Distribution:} We present the distribution of the emotion labels obtained from participant responses. Our data is equitably distributed among the four emotion categories.}
  \vspace{-20pt}
  \label{fig:dataDistribution}
\end{figure}

We present the distribution of the gaits and their associated emotions in Figure~\ref{fig:dataDistribution}. The smallest emotion category, \textit{Neutral}, contains $16.35\%$ of the gaits, whereas the largest emotion category, \textit{Happy}, contains $32.07\%$, indicating that our data is equitably distributed among the four emotion categories.

\subsection{Application: Prediction of Perceived Emotion}\label{sec:gaitClassification}

We present an application of our gait emotion association to predict the perceived emotion given any new gait. This allows us to extend our gait emotion association to include any new gait by predicting its perceived emotion value without having to obtain ratings from participants. 

\begin{figure}[t]
  \centering
  \includegraphics[width =\linewidth]{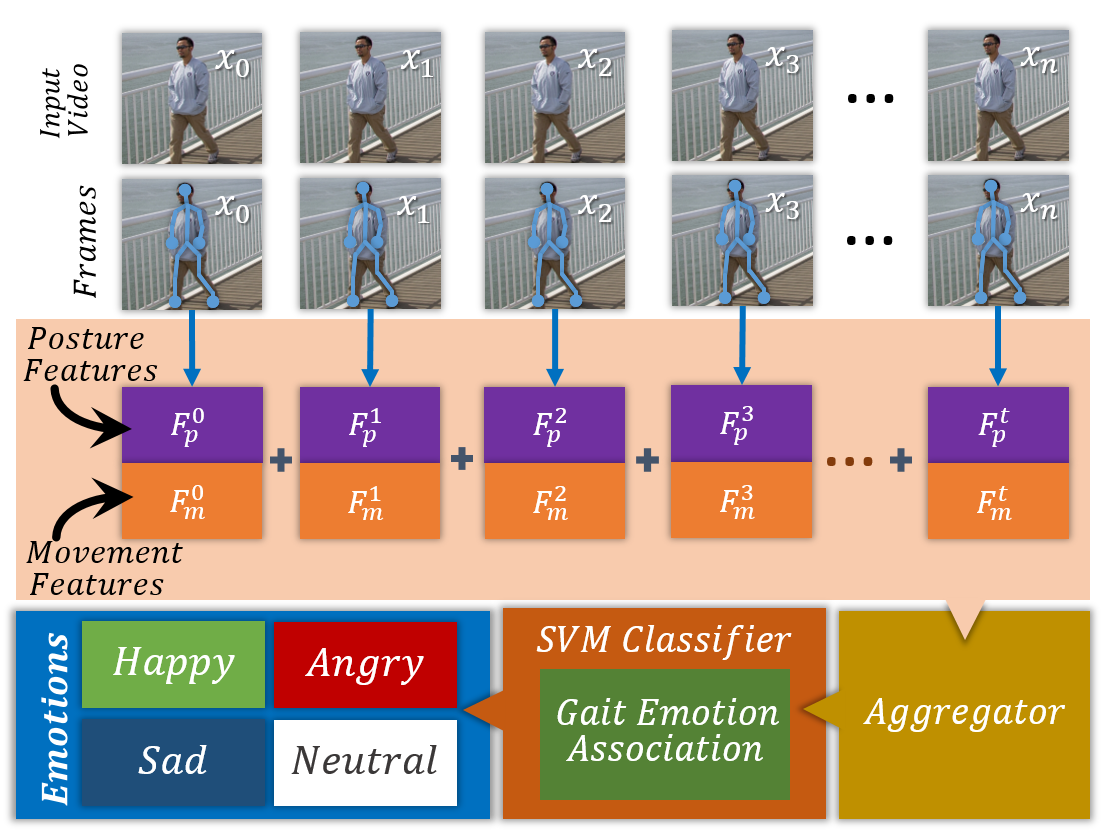}
  \vspace{-15pt}
  \caption{\textbf{Prediction of Perceived Emotion:} We can use gait emotion association to predict perceived emotions of any new gait. Given a new gait, we compute gait features using psychological characterization~\protect\cite{crenn2016body}. We use an SVM classifier trained using the gait emotion association to classify these features into one of the four basic emotions with $70.04\%$ accuracy.}
  \vspace{-18pt}
  \label{fig:classificationOverview}
\end{figure}

We provide an overview of the application framework in Figure~\ref{fig:classificationOverview}. We use the gait emotion association to train an SVM.
Given a gait, we compute numerical values for its expressive features using psychological characteristics~\cite{crenn2016body,karg2010recognition} corresponding to posture and movement of joints. We use the SVM to classify our computed gait features into one of the four basic emotions. We extract features corresponding to both posture and movement modalities because both are important for the accurate prediction of an individual's emotional state~\cite{kleinsmith2013affective}. We assume a skeleton with $16$ joints. Before extracting features, we extract frames corresponding to a single stride from the video, i.e., frames corresponding to consecutive foot strikes of the same foot (walk cycle). We compute the posture and movement features at each frame and aggregate them over the walk cycle.

\subsubsection{Posture Features}
We use the following posture features:\\
$\bullet$ \textbf{Volume}: According to Crenn et al.~\cite{crenn2016body}, body expansion conveys positive emotions, whereas a person has a more compact posture during negative expressions. We model this by a single scalar value corresponding to the volume occupied by the bounding box around the human.\\
$\bullet$ \textbf{Area}: We model body expansion by two scalar values corresponding to the areas of triangles between the hands and the neck and between the feet and the root joint.\\
$\bullet$ \textbf{Distance}: Distances of the feet and the hands from the root joint can be used to model body expansion (four scalar values of distances).\\
$\bullet$ \textbf{Angle}: Head tilt is used to distinguish between happy and sad emotions~\cite{crenn2016body,karg2010recognition}. We model this using five scalar values corresponding to the angles: (1) at the neck by the shoulder joints, (2) at the shoulder joints by the neck and the other shoulder, (3) at the neck by the vertical direction and the back, and (4) at the neck by the head and the back.

We also include stride length as a posture feature. Longer stride lengths convey anger and happiness, and shorter stride lengths convey sadness and neutrality~\cite{karg2010recognition}. Suppose we represent the positions of the left foot joint $j_{lFoot}$ and the right foot joint $j_{rFoot}$ in frame $t$ as $\vec{p}(j_{lFoot}, t)$ and $\vec{p}(j_{rFoot}, t)$, respectively. Then the stride length $s \in \mathbb{R}$ is computed as:
\begin{eqnarray}
    s = \max\limits_{t \in 1..\tau}||\vec{p}(j_{lFoot}, t) - \vec{p}(j_{rFoot}, t)||.
\end{eqnarray}

We represent the posture features at frame $t=\{1,2,..,\tau\}$ by a 13-dimensional vector $F_{p, t}\in \mathbb{R}^{13}$, where $\tau$ is the number of frames in the walk cycle. Then, we can compute the posture features of the gait $F_{p}$ as the average of the posture features at each frame $t$ combined with the stride length:
\begin{eqnarray}
    F_{p} = \frac{\sum_{t} F_{p, t}}{\tau} \cup s.
\end{eqnarray}

\subsubsection{Movement Features}
High arousal emotions such as anger and happiness are associated with rapid and increased movements, whereas low arousal emotions such as sadness are associated with slow movements~\cite{bernhardt2007detecting}. We consider the magnitude of the velocity, acceleration, and movement jerk of the hand, foot, and head joints and compute the movement features $F_{m, t} \in \mathbb{R}^{15}$ at frame $t$. For each of these five joints $j_i, i={1,...,5}$, we compute the magnitude of the first, second, and third derivatives of the position vector $\vec{p}(j_i, t)$ at frame $t$. 

Since faster gaits are perceived as happy or angry and slower gaits are considered as sad~\cite{karg2010recognition}, we also include the time taken for one walk cycle ($gt\in \mathbb{R}$) as a movement feature. We define the movement features $F_{m}\in \mathbb{R}^{16}$ as the average of $F_{m, t}, t=\{1,2,..,\tau\}$:
\begin{eqnarray}
    F_{m} = \frac{\sum_{t} F_{m, t}}{\tau} \cup gt.
\end{eqnarray}

We combine posture and movement features and compute the gait features $F$ as $F = F_{m} \cup  F_{p}$. In the end, we normalize the feature values to $[-1, 1]$ with $-1$ and $1$ representing the minimum and maximum values of the feature in the dataset, respectively.

\subsubsection{Emotion Classification Accuracy}
For each gait $\textbf{G}_i$ in the dataset, we have computed the gait features $F_{i}$ and the associated perceived emotion label $e_i$. Using this gait emotion association, we train an SVM classifier with an RBF kernel with a  one-versus-rest decision function of shape.  For any new gait, we compute the values of the gait features and use the trained SVM to predict its perceived emotion. We obtain the accuracy results by performing 10-fold cross-validation on all six datasets (Section~\ref{sec:datasets}). We compare the accuracy (Table~\ref{tab:accuracySota}) of our classification algorithm with other state-of-the-art methods that predict emotions based on gaits:
\begin{itemize}
    \setlength\itemsep{0pt}
    \item Karg et al.~\shortcite{karg2010recognition}:  This feature-based classification method uses PCA to classify gait features related to shoulder, neck, and thorax angles, stride length, and velocity. This method only models the posture features for the joints and doesn't model the movement features.
    \item Venture et al.~\shortcite{venture2014recognizing}: This method uses the auto-correlation matrix of the joint angles at each frame and uses similarity indices for classification. Their algorithm achieves good intra-subject accuracy but performs poorly for inter-subject databases.
    \item Crenn et al.~\shortcite{crenn2016body}: This feature-based method uses both posture and movement features and classifies these features using SVMs. This method considers only the upper body and does not model features related to feet joints for classification. 
    \item Daoudi et al.~\shortcite{daoudi2017emotion}: This method uses a manifold of symmetric positive definite matrices to represent body movement and classifies them using the nearest neighbors method.
    \item Crenn et al.~\shortcite{crenn2017toward}: This method is based on the synthesis of neutral motion from an input motion and uses the residual between the input and neutral motion for classification.
\end{itemize}

\begin{table}[]
\centering
\begin{tabular}{|l|l|}
\hline
\multicolumn{1}{|c|}{\textbf{Method}}         & \textbf{Accuracy} \\ \hline
Karg et al.~\cite{karg2010recognition}    & 39.58\%         \\ \hline
Venture et al.~\cite{venture2014recognizing} &    30.83\%      \\ \hline
Crenn et al.~\cite{crenn2016body}   &   66.22\%       \\ \hline
Crenn et al.~\cite{crenn2017toward}   &  40.63\%       \\ \hline
Daoudi et al.~\cite{daoudi2017emotion}  &   42.52\%       \\ \hline
\textit{Our Method (PEP)}    &   \textbf{70.04\%}       \\ \hline
\end{tabular}
\caption{\textbf{Accuracy}: Our method using gait features, including posture and movement features, and SVM classifier achieves an accuracy of $70.04\%$ for emotion classification. We obtain considerable improvement over prior methods.}
\vspace{-25pt}
\label{tab:accuracySota}
\end{table}





\section{Emotions in Virtual Agents (EVA)}
In this section, we provide the details of our \textit{EVA} algorithm, which uses the gait emotion association to generate emotions in virtual agents.

\subsection{Notation}
We define a gait $\textbf{G}$ as a set of 3D poses ${P_1, P_2,..., P_{\tau}}$ where $\tau$ is the number of frames. The set of 3D positions of each joint $J_i$ is referred to as the pose of an agent $P \in \mathbb{R}^{3*n}$ where $n$ is the number of joints.

\subsection{Gait Generation}\label{sec:gaitGeneration}
At runtime, for each agent, we take the desired emotion $e_{des}$ as input. We use the gait emotion association metric and obtain a set of gaits $GEA(e_{des})$ that correspond to the desired emotion. We choose one of the gaits $G_{des}$ from $GEA(e_{des})$ and update the joint positions of the agent in the virtual world using the joint positions from $G_{des}$. The selection of  $G_{des}$ from $GEA(e_{des})$ can be made according to many criteria (such as personality or preferred walking speed). In our current implementation, we choose the gait randomly.

\subsection{Local Navigation using Expressive Features}
For navigation, we represent each agent on the 2D ground plane and compute its collision-free velocity. We use a reciprocal velocity obstacle-based method, RVO, to generate smooth, stable, collision-free velocities~\cite{van2008reciprocal}. RVO is an agent-based approach that computes a collision-free 2D velocity of an agent given its preferred velocity, time horizon ($t_{max}$), and current positions and velocities of the all virtual agents in the environment. In other words, it computes a velocity that can generate a collision-free trajectory at time $t_{max}$.

Based on the chosen gait $G_{des}$ for an agent $i$, we set the preferred speed of the agent as the speed of the gait $G_{des}$. The speed of the gait $v_{des}$ is computed as follows:
\begin{eqnarray}
    v_{des} = ||\vec{p(J_{root,\tau})} - \vec{p(J_{root,0})}||/\tau,
\end{eqnarray}
where $\vec{p(J_{root,\tau})}$ and $\vec{p(J_{root,0})}$ represent the positions of the root joint in the first $P_0$ and last pose $P_{\tau}$, respectively.

\subsection{Gaze Behaviors}
According to psychology literature, the direction of the gaze affects the perception of emotions~\cite{adams2003perceived}. Direct gazes are associated with approach-oriented emotions such as anger and averted gazes are associated with avoidance-oriented emotions such as sadness~\cite{adams2003perceived}. Using these observations, we present an emotion-based gaze formulation. 

\begin{figure}[t]
  \centering
  \includegraphics[width =\linewidth]{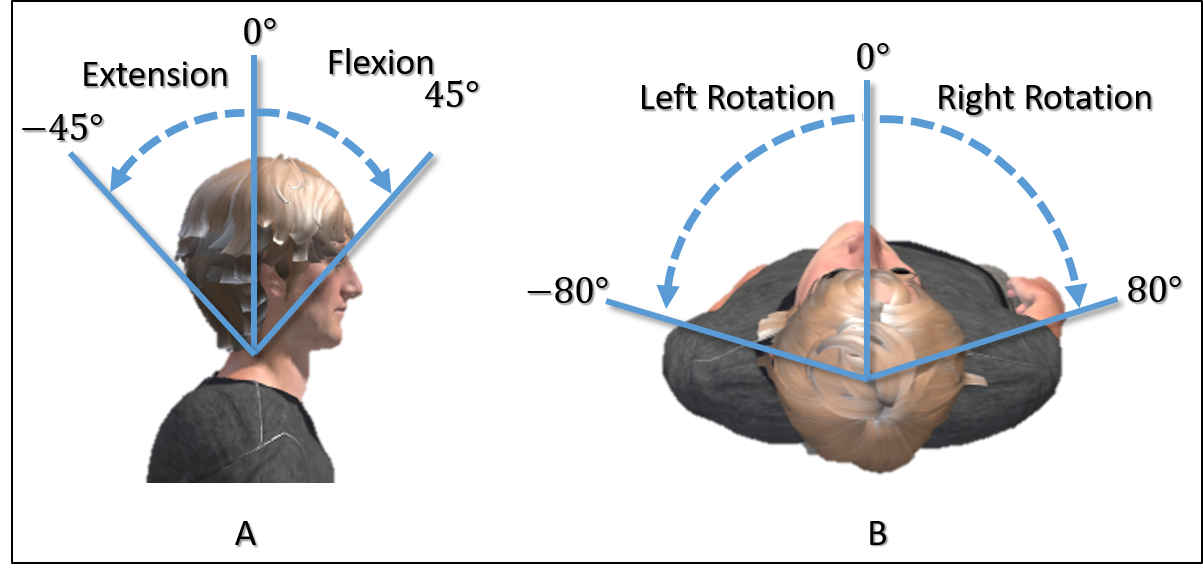}
  \vspace{-18pt}
  \caption{\textbf{Gaze Control:} We control virtual agents' gaze by controlling the rotations of the neck joint in the skeleton mesh. Specifically, we control the two degrees of freedoms associated with (1) neck flexion and extension, and (2) neck left and right rotation.}
  \vspace{-18pt}
  \label{fig:neckAngles}
\end{figure}

We control a virtual agent's gaze by controlling the rotations of the neck joint in the skeleton mesh. Specifically, we control the angles corresponding to two degrees-of-freedom: (1) neck flexion and extension and (2) neck left and right rotation. Given the desired emotion $e_{des}$ for a virtual agent $i$, our goal to compute the two angles $\theta_{flex}$ and $\theta_{rot}$. We present a visualization of our gaze control angles in Figure~\ref{fig:neckAngles}. Here, the subscripts $i$ and $u$ represent the virtual agent and the user respectively, and $(x, y, z)$ represent the 3D coordinates.

\noindent $\bullet$ \textbf{Happy}: According to prior literature, direct gaze and neck extension are associated with happiness~\cite{adams2003perceived}. Therefore, for $e_{des} = happy $, the agent's neck orientation is computed such that the virtual agent gazes directly at the user with an upward head tilt. The neck angles corresponding to $e_{des} = happy $ are computed as follows:
\begin{eqnarray}
    \theta_{flex} &=& \theta_{happy},  \\
    \theta_{rot} &=& angle(\vec{p_u}-\vec{p_i}, \vec{f}),
\end{eqnarray}
where $\theta_{happy}$ is a negative constant value indicating neck extension and $angle(*,*)$ represents the angle between two vectors. We use $\theta_{happy} = -5^{\circ}$ for our validation study.

\noindent $\bullet$ \textbf{Angry}: Direct gaze is associated with approach-based emotions such as anger~\cite{adams2003perceived}. Therefore, for $e_{des} = angry$, the agent's neck orientation is computed such that the agent gazes directly at the user agent. The neck angles for $e_{des} = angry$ are computed as follows:
\begin{eqnarray}
    \theta_{flex} &=& \arcsin (\frac{y_i-y_u}{\sqrt{(x_i - x_u)^2 + (z_i - z_u)^2}}), \\
    \theta_{rot} &=& angle(\vec{p_u}-\vec{p_i}, \vec{f}),
\end{eqnarray}
where $angle(*,*)$ represents the angle between two vectors.

\noindent $\bullet$ \textbf{Sad}: Averted gaze and neck flexion are associated with sadness~\cite{adams2003perceived}. The neck angles for $e_{des} = sad$ are computed as follows:
\begin{eqnarray}
    \theta_{flex} &=& \theta_{sad}, \\
    \theta_{rot} &=& 0^{\circ},
\end{eqnarray}
where $\theta_{sad}$ is a negative constant value indicating neck extension. We use $\theta_{sad} = 10^{\circ}$ for our validation study.

\noindent $\bullet$ \textbf{Neutral}: For the neutral emotion, we use the orientations of the neck angles $\theta_{flex}$ and $\theta_{rot}$ computed from the positions of the head and neck joints in the gait $G_{des}$.

\subsection{EVA Algorithm} 
During runtime, we first compute desired gait $G_{des}$ using gait generation method described in Section~\ref{sec:gaitGeneration}. Next, we compute collision-free trajectories and update the position of the agents in the ground plane. We retarget the generated gait $G_{des}$ to the avatar of the virtual agent using Unity's skeleton retargeting~\cite{skeletonRetargeting} and update the positions of the agent's joints in the virtual world using the joint positions from $G_{des}$. We use the emotion-based gazing behaviors to update the neck joint's flexion and rotation angles. The resultant virtual agents are then rendered to an HMD.

\section{User Evaluation}
In this section, we present the details of our user evaluation conducted to evaluate the benefits of our virtual agent movement generation algorithm \textit{(EVA)} with and without gazing behaviors. We compare its performance with a baseline method.

\subsection{Experiment Goals and Expectations}
We performed the user evaluation to determine whether or not significant improvement in the sense of presence can be observed by a real user immersed in a virtual environment based on the emotional state of virtual agents computed using EVA. We chose the sense of presence metric because it has been widely used to evaluate the quality of a VR experience, and a higher sense of presence is desirable for most VR applications~\cite{pedvr,garau2005responses}. We wanted to demonstrate that our algorithm improves the sense of presence, thus improving the overall quality of the simulations. We compared the following three virtual agent generation algorithms:

\noindent $\bullet$ \textbf{Baseline}:  All virtual agents were simulated with a single neutral gait. Virtual agents did not perform any gazing behaviors. 

\noindent $\bullet$  \textbf{EVA-O}:  Our novel virtual agent generation algorithm \textit{(EVA)} without any gazing behaviors. We generate gaits corresponding to the four emotions: \textit{happy, angry, sad}, or \textit{neutral}.

\noindent $\bullet$  \textbf{EVA-G}: Our novel virtual agent generation algorithm \textit{(EVA)} with emotion-specific gaits along with the gazing behaviors.

Additionally, we also performed an emotion identification task, where we wanted to estimate the accuracy of our EVA algorithm in generating characters with the desired perceived emotions.

\subsection{Experimental Design}
We conducted the study using a within-subjects design. We showed each participant four scenarios with virtual agents walking in different environments. For each scenario, participants performed three trials corresponding to the three different virtual agent generation algorithms. The order of the scenarios and the trials was counterbalanced. In each of the scenarios, we used a set of male and female virtual agents. For \textit{EVA-O} and \textit{EVA-G}, the virtual agents were simulated with desired emotions chosen randomly out of the four categories: \textit{Happy, Angry, Sad,} and \textit{Neutral}. For the baseline, all the agents were simulated using the gaits corresponding to the \textit{Neutral} category.

After the four scenarios were completed, the participants performed the emotion identification task. For this task, we generated eight virtual agents: two agents corresponding to each desired emotion from \textit{Happy, Angry, Sad,} and \textit{Neutral}. We assigned male and female virtual agents to each emotion in this task as well. We displayed these virtual agents one-by-one to the participants. Participants were asked to classify these agents as \textit{Happy, Angry, Sad,} and \textit{Neutral}. 


\begin{figure*}[t]
    \centering
    \includegraphics[width=1.0\linewidth, height=200px]{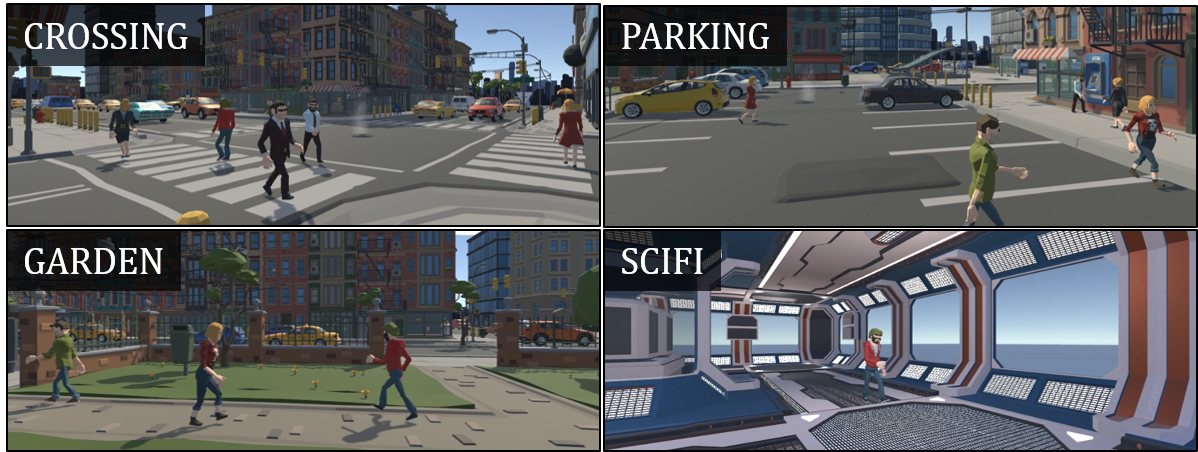}
    \vspace{-18pt}
    \caption{\textbf{Scenarios:} Our user evaluation consists of these $4$ scenarios.}
    \vspace{-15pt}
    \label{fig:scenarios}
\end{figure*}


\subsubsection{Procedure}
We welcomed the participants and informed them about the overall process and purpose of the study. We introduced them to an HTC Vive HMD and informed that it might cause nausea and slight discomfort. The experiment was approved by Institutional Review Boards and the Office of Human Research Ethics. Before beginning the experiment, the participants were invited to read and agree to the approved consent form. They were also informed that they could opt out of the experiment at any time. During the study, the participants could walk in the tracking space and look around in the virtual environment by rotating their heads. Participants provided optional demographic information about their gender and age. The study required approximately $20$ minutes, and participants were compensated with a gift card worth \$$5$.

\subsubsection{Scenarios}
We evaluated the virtual agent generation algorithms in $4$ different scenarios (Figure~\ref{fig:scenarios}):

\begin{itemize}
    \setlength\itemsep{0pt}
    \item \textbf{Crossing}: This scenario had $8$ virtual agents crossing a busy intersection in a city environment.
    \item \textbf{Garden}: This scenario had $4$ virtual agents walking in a small leisure garden.
    \item \textbf{Parking}: This scenario consisted of $8$ virtual agents walking in a parking lot near a set of stores.
    \item \textbf{Scifi}: This scenario consisted of $4$ virtual agents walking inside a fantastical residential pod.
\end{itemize}

In each of the scenario, we chose characters with low polygon counts because they provide the capability to simulate a large number of virtual agents while maintaining the necessary frame rate for VR. It also avoids the uncanny valley problem~\cite{geller2008overcoming}.




\subsubsection{Participants}
A total of $30$ participants ($20$ male, $10$ female,  $\bar{x}_{age}$ = $25.4$) from a university took part in the study.

\subsubsection{Questions} 
To test the proposed hypotheses, we used a modified version of a well-established questionnaire by Garau et al~\shortcite{garau2005responses}. These questions attempt to assess the various aspects of social presence~\cite{lombard2009measuring}. In addition to a subset of the original questions, we asked a question whether the virtual agents seemed to be experiencing different emotions. Participants answered the Agree/Disagree type questions on seven-level Likert items. We ask the following questions:\\
\textbf{Q1}, \textit{Spatial Presence}: I had a sense of being in the same space as the characters. \\
\textbf{Q2}, \textit{Awareness}: The characters seemed to be aware of me. \\
\textbf{Q3}, \textit{Interaction}: I felt that I should talk/nod/respond to the characters. \\
\textbf{Q4}, \textit{Realism}: The characters seemed to resemble real people. \\
\textbf{Q5}, \textit{Emotions}: The characters seemed to be experiencing different emotions.

\subsection{Results} 
\begin{figure}[t]
  \centering
  \includegraphics[width =\linewidth]{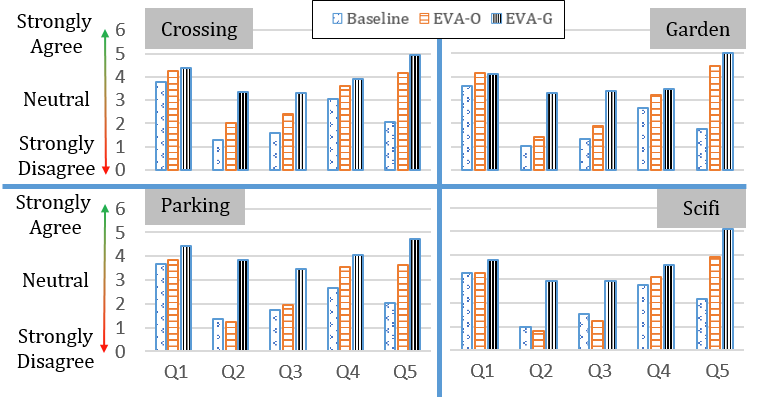}
  \vspace{-15pt}
  \caption{\textbf{Averaged Responses}: We present the average participant responses for all scenarios. These results indicate that the participants consistently preferred our \textit{EVA} algorithm over the baseline method.}
  \vspace{-15pt}
  \label{fig:means}
\end{figure}

In this section, we discuss the participants' responses. We computed average participant responses for all scenarios (Figure~\ref{fig:means}). We also tested the differences between responses for the three algorithms using the \textit{Friedman test}. For this test, the simulation algorithm is the independent variable, and the participant response is the dependent variable. We present the test statistic value $\chi^2$ and the significance level $p$ in Table~\ref{tab:friedman}. We observe significant differences between the three compared methods for all questions across all scenarios (except Q1 in Scifi and Garden scenario). We also performed a post hoc analysis using the \textit{Wilcoxon Signed-Rank tests}. We applied Bonferroni correction resulting in a significance level at $p < 0.017$.  We present the significance level $(p)$ for pairwise comparisons between the three algorithms in Table~\ref{tab:wilcoxon}. We observe significant differences between baseline and EVA-G, indicating that our method of generating emotions in virtual agents using both gait and gaze features performs consistently better than the baseline method. We discuss the results in detail below:

\begin{table}[t]
\centering
\resizebox{\linewidth}{!}{
\begin{tabular}{|c|c|c|c|c|c|c|c|c|}
\hline
\multirow{2}{*}{} & \multicolumn{2}{c|}{Crossing} & \multicolumn{2}{c|}{Garden} & \multicolumn{2}{c|}{Parking} & \multicolumn{2}{c|}{Scifi} \\ \cline{2-9} 
                   & $\chi^2$      & p         & $\chi^2$          & p             & $\chi^2$          & p             & $\chi^2$            & p              \\ \hline
Q1                & 9.75           & 0.01         & 4.67          & 0.10        & 15.13         & 0.00         & 2.46         & 0.29        \\ \hline
Q2                & 19.58          & 0.00         & 26.45         & 0.00        & 30.28         & 0.00         & 27.30        & 0.00        \\ \hline
Q3                & 20.26          & 0.00         & 27.27         & 0.00        & 20.17         & 0.00         & 24.90        & 0.00        \\ \hline
Q4                & 13.77          & 0.00         & 12.30         & 0.00        & 24.45         & 0.00         & 10.53        & 0.01        \\ \hline
Q5                & 23.36          & 0.00         & 41.32         & 0.00        & 35.04         & 0.00         & 43.38        & 0.00        \\ \hline
\end{tabular}}
\caption{\textbf{Results of the Friedman Test}: We present the test statistic $(\chi^2)$ and the significance level $(p)$ for each scenario for all questions. The values of $p < 0.05$ indicate a significant difference between the responses for the three algorithms. }
  \vspace{-18pt}
\label{tab:friedman}
\end{table}

\begin{table}[t]
\begin{tabular}{|cl|c|c|c|c|c|}
\hline
\multicolumn{3}{|c|}{\multirow{2}{*}{}}                       & \multicolumn{4}{c|}{Participant Response} \\ \cline{4-7} 
\multicolumn{3}{|c|}{}                                        & Happy    & Angry   & Sad      & Neutral   \\ \hline
\multirow{4}{*}{\rotatebox{90}{Desired}} & \multirow{4}{*}{\rotatebox{90}{Emotion}} & Happy   & 60.41\%    & 4.17 \%   & 4.17  \%   & 31.25\%     \\ \cline{3-7} 
                         &                          & Angry   & 18.75\%    & 52.08\%   & 0.00  \%   & 29.17\%     \\ \cline{3-7} 
                         &                          & Sad     & 0.00 \%    & 0.00 \%   & 100.00\%   & 0.00 \%     \\ \cline{3-7} 
                         &                          & Neutral & 8.33 \%    & 4.17 \%   & 16.67 \%   & 70.83\%     \\ \hline
\end{tabular}
\caption{\textbf{Confusion Matrix}: We present the confusion matrix for the emotion identification task. Here, rows indicate the desired emotion input provided to the EVA algorithm, and the columns indicate the participants' responses.}
\label{tab:confusion}
  \vspace{-18pt}
\end{table}

\begin{table*}[!]
\resizebox{\linewidth}{!}{
\begin{tabular}{|c|c|c|c|c|c|c|c|c|c|c|c|c|}
\hline
\multirow{2}{*}{} & \multicolumn{3}{c|}{Crossing}                                                                                                                                                              & \multicolumn{3}{c|}{Garden}                                                                                                                                                                 & \multicolumn{3}{c|}{Parking}                                                                                                                                                                 & \multicolumn{3}{c|}{Scifi}                                                                                                                                                                  \\ \cline{2-13} 
                  & \begin{tabular}[c]{@{}c@{}}Baseline\\ vs \\ EVA-O\end{tabular} & \begin{tabular}[c]{@{}c@{}}EVA-O \\ vs \\ EVA-G\end{tabular} & \begin{tabular}[c]{@{}c@{}}Baseline\\ vs\\  EVA-G\end{tabular} & \begin{tabular}[c]{@{}c@{}}Baseline \\ vs\\ EVA-O\end{tabular} & \begin{tabular}[c]{@{}c@{}}EVA-O \\ vs \\ EVA-G\end{tabular} & \begin{tabular}[c]{@{}c@{}}Baseline\\  vs \\ EVA-G\end{tabular} & \begin{tabular}[c]{@{}c@{}}Baseline\\  vs \\ EVA-O\end{tabular} & \begin{tabular}[c]{@{}c@{}}EVA-O\\  vs \\ EVA-G\end{tabular} & \begin{tabular}[c]{@{}c@{}}Baseline\\  vs \\ EVA-G\end{tabular} & \begin{tabular}[c]{@{}c@{}}Baseline\\  vs \\ EVA-O\end{tabular} & \begin{tabular}[c]{@{}c@{}}EVA-O\\  vs \\ EVA-G\end{tabular} & \begin{tabular}[c]{@{}c@{}}Baseline\\ vs \\ EVA-G\end{tabular} \\ \hline
Q1                & 0.018                                                          & 0.323                                                      & 0.005                                                        & NA                                                             & NA                                                         & NA                                                            & 0.179                                                           & 0.019                                                      & 0.003                                                         & NA                                                              & NA                                                         & NA                                                           \\ \hline
Q2                & 0.037                                                          & 0.002                                                      & 0.000                                                        & 0.095                                                          & 0.000                                                      & 0.000                                                         & 0.540                                                           & 0.000                                                      & 0.000                                                         & 0.179                                                           & 0.000                                                      & 0.000                                                        \\ \hline
Q3                & 0.012                                                          & 0.005                                                      & 0.000                                                        & 0.007                                                          & 0.000                                                      & 0.000                                                         & 0.256                                                           & 0.001                                                      & 0.000                                                         & 0.153                                                           & 0.000                                                      & 0.000                                                        \\ \hline
Q4                & 0.038                                                          & 0.276                                                      & 0.001                                                        & 0.026                                                          & 0.160                                                      & 0.012                                                         & 0.005                                                           & 0.035                                                      & 0.000                                                         & 0.054                                                           & 0.041                                                      & 0.010                                                        \\ \hline
Q5                & 0.000                                                          & 0.011                                                      & 0.000                                                        & 0.000                                                          & 0.070                                                      & 0.000                                                         & 0.001                                                           & 0.000                                                      & 0.000                                                         & 0.000                                                           & 0.000                                                      & 0.000                                                        \\ \hline
\end{tabular}}
\caption{\textbf{Results of the Wilcoxon Signed-Rank Test}: We present the significance level $(p)$ for pairwise comparisons between the three algorithms using the post hoc tests. The values of $p < 0.017$ indicate a significant difference between the responses for corresponding pairwise comparisons.  }
  \vspace{-18pt}
\label{tab:wilcoxon}
\end{table*}

\noindent $\bullet$ \textbf{Spatial Presence}: According to the higher averaged responses, participants experienced a higher sense of being in the same space as the characters simulated with EVA, as compared to the baseline. For Garden and Scifi scenarios, we do not observe a significant difference between the sense of presence felt by the participants across the three algorithms according to the Friedman test. Because of the insignificant differences in the Friedman test, post hoc tests do not apply to these scenarios. For Crossing and Parking scenarios, where there were more agents, we observe a significant difference between the responses for different methods in the Friedman test. The Wilcoxon tests reveal significant differences between the \textit{baseline} and \textit{EVA-G} methods. 

\noindent $\bullet$ \textbf{Awareness}: Participants felt that the virtual agents were more aware of them compared to the baseline method as evidenced by the averaged responses. We observe significant differences between the responses across all scenarios using the Friedman test. Wilcoxon tests reveal significant differences in the \textit{baseline vs. EVA-G} and \textit{EVA-O vs. EVA-G} comparisons, highlighting the importance of gazing behaviors in our algorithm.

\noindent $\bullet$ \textbf{Interaction}: Participants felt that they should talk/nod/respond to the agents if their movement was simulated by EVA as compared to the baseline as evidenced by the averaged responses. We observe significant differences between the responses across all scenarios using the Friedman test. Wilcoxon tests reveal significant differences for the \textit{baseline vs. EVA-G} and \textit{EVA-O vs. EVA-G} comparisons highlighting the importance of gazing behaviors in our algorithm. We also observe a significant difference for the \textit{baseline vs. EVA-O} comparisons for the Crossing and Garden scenarios, indicating the importance of gaits in eliciting more response from the users.

\noindent $\bullet$ \textbf{Realism}: When EVA simulated the agents, they seemed to resemble real people more than when they were simulated by the baseline algorithm as indicated by the participant responses. We observe significant differences between the responses across all scenarios using the Friedman test. Wilcoxon tests reveal significant differences in the \textit{baseline vs. EVA-G}, highlighting the importance of both gait and gazing behaviors in our algorithm.

\noindent $\bullet$ \textbf{Emotions}: The agents seemed more likely to be experiencing different emotions if the EVA algorithm simulated them compared to the baseline algorithm according to the averaged responses. We observe significant differences between the responses across all scenarios using the Friedman test. Wilcoxon tests reveal significant differences for all the pairwise comparisons (except for \textit{EVA-G vs. EVA} comparison in the Garden scenario).

\subsection{Benefits of Expressive Features}
These results show that expressive features provide the following benefits:

\noindent $\bullet$  for scenarios with a higher number of virtual agents, virtual agents simulated using expressive features improve the sense of presence felt by the user,

\noindent $\bullet$  expressive features make the virtual agents appear more aware of the user,
 
\noindent $\bullet$  elicit more response from the users, and

\noindent $\bullet$ increase the resemblance of the virtual agents to real people.

\subsection{Emotion Identification}
For the emotion identification task, the participants identified the emotions of the virtual agents correctly with an accuracy of $70.83\%$. We present the confusion matrix in Table~\ref{tab:confusion}. We obtain $100\%$ accuracy in simulating virtual agents that are perceived as sad and more than $50\%$ accuracy for all the emotions. These results indicate that expressive features provide benefits for the perception of emotions in virtual agents.

\section{Conclusions, Limitations, and Future Work}
We present a data-driven approach for generating emotions in virtual agents \textit{(EVA)} using expressive features. We use a novel data-driven metric between gaits and emotions. We combine these gaits with gazing behaviors and generate virtual agents that are perceived as experiencing \textit{happy, angry, sad}, or \textit{neutral} emotions. We also present an application of our gait emotion association to predict how the real users will perceive the emotions of virtual agents simulated with any new gait and achieve an accuracy of $70.04\%$. We validated our EVA algorithm with user evaluation. 
Results of this evaluation indicate that the expressive features can increase the sense of presence felt by the user in the virtual environment and make the virtual agents elicit more response from them. 

Our approach has some limitations. Our approach only takes into account expressive features of gait and gaze, and ignore the impact of other agent characteristics related to appearance or facial expressions or other non-verbal cues like appearance, voice, pupil dilation, etc. for the perception of emotion~\cite{kret2013emotional}. We conducted the evaluation in a multi-agent VR environment to measure the contribution of our algorithm in a more interactive and immersive setting. In the future, we would like to examine the gait emotion association metric per agent to strengthen the desired emotion as much as possible while dampening the others. Furthermore, we limited this approach to walking activity. In future work, we would like to extend our approach to other activities. Since previous research shows that gender is recognizable from gaits~\cite{zibrek2015exploring}, we would like to consider the gender of the virtual agents while generating a gait for them. We would also like to combine our approach with other emotional cues and compare it with different algorithms that simulate other types of motion styles (e.g., personalities~\cite{durupinar2017perform}). Finally, we would like to examine the impact of the context and culture of the observer on the perception of emotion from bodily expressions using recent psychological findings~\cite{gelder_2016}.

\bibliographystyle{ACM-Reference-Format}
\bibliography{template}

\end{document}